\documentclass{aa}
\usepackage{graphicx} 
\usepackage{natbib} 
\def\gc{\object{$\gamma$\,Cephei}} 
\begin{document} 
\title{Planetary formation in the $\gamma$ Cephei system} 
 
\author{P. Th\'ebault\inst{1}, F. Marzari\inst{2}, 
H. Scholl\inst{3}, D. Turrini\inst{2},  M. Barbieri\inst{1,2}} 
\institute{ Observatoire de Paris, Section de Meudon, 
F-92195 Meudon Principal Cedex, France 
\and 
Dipartimento di Fisica, Universita di Padova, Via Marzolo 8, I-35131 
Padova, Italy 
\and 
Observatoire de la C\^ote d'Azur, Dept. Cassiop\'ee, B.P. 4229, 
F-06304 Nice, France} 
\offprints{P. Th\'ebault} \mail{philippe.thebault@obspm.fr} 
\date{Received; accepted} \titlerunning{The $\gamma$ Cephei System} 
\authorrunning{Th\'ebault et al.} 
 
\abstract { 
We numerically investigate under which conditions the 
planet detected at 2.1 AU of \gc\ could form through 
the core-accretion scenario despite the perturbing presence 
of the highly eccentric companion star. 
We first show that the initial stage of runaway accretion 
of kilometer-sized planetesimals is possible within 2.5 
AU from the central star only if 
large amounts of gas are present. In this case, gaseous 
friction induces periastron alignment of the orbits which 
reduces the otherwise high mutual impact velocities due 
to the companion's secular perturbations. 
The following stage of mutual accretion of large embryos 
is also modeled. According to our 
simulations, the giant impacts among the embryos 
always lead to a core of 10 $M_{\oplus}$ within 
10 Myr, the average lifetime of gaseous discs. However, 
the core always ends up within 1.5 AU from the central 
star. Either the core grows more quickly in the inner region 
of the disc, or it migrates inside by scattering the 
residual embryos. 
\keywords{stars: planetary systems -- 
        planetary systems: formation -- 
        planets and satellites: formation 
                       
                       } 
                          } 
 
\maketitle 
 
% 
%________________________________________________________________ 

\section{Introduction} 
 
Among the 15 presently known  binary star systems harbouring 
extra solar planets, \gc\ is that with the closest companion 
star \citep[with the 
exception of Gliese 86, where the possible companion is believed to be 
a brown dwarf][]{els01}. 
According to \citet{hatz03}, the secondary star has an orbit with a semimajor 
axis of $18.5 \pm 1.1$ AU and an eccentricity of $0.361 \pm 0.023$. 
The planet detected around the primary K0 III giant star, 
has a mass of $M sin ~ i = 1.7 \pm 0.4$ Jupiter masses and an orbital 
semimajor axis of 2.13 AU. The planet is located 
inside a stable region with a dynamical 
lifetime of at least 1 Gyr. This region extends to about 4 AU 
from the central star \citep{how99,dvo03}. 
 
Giant planets orbiting one of the stars in a 
binary system offer the possibility to test the core--accretion model 
for giant planet formation \citep{pol96} in a complex 
dynamical environment. The vicinity of a companion star 
on a highly eccentric orbit may prevent planetary formation 
because the companion reduces the size of the accretion disc, and 
 it excites high relative velocities 
between colliding planetesimals. 
 
In this paper we study the {\it in situ} formation of the giant 
planet in the \gc\ system by numerically investigating 
both the stage of planetary embryo accretion from planetesimals 
and the following stage 
of high velocity collisions between large embryos 
which form the core of the giant planet. 
This last stage precedes that of rapid gas accretion which 
sets on after sufficient accumulation of mass onto the 
core. 
 
A critical parameter for the initial stage of 
planetesimal accretion is the 
{\it relative impact velocity} $\Delta$v. This velocity 
determines whether accretion or erosion 
dominates the planetesimal collisional 
evolution. In a binary star system, the 
secular perturbations of the closeby companion star may play 
a critical role by exciting the relative velocities among the 
planetesimals. Too large $\Delta$v prevent planetesimals 
from growing by accretion and 
the initial planetesimal population may even be ground down to dust. 
We explored the distribution  of collisional relative velocities of a 
planetesimal  population surrounding the primary star of the \gc\ system 
by using a deterministic code that includes the effect of gas drag. 
This later mechanism might indeed play a crucial role since 
gas drag might force periastron and eccentricity 
alignment of the perturbed planetesimals orbits and thus 
reduce the large $\Delta$v induced by the companion's perturbations 
\citep[e.g.][]{mascho00}. 
 
We also investigated the late stage of core formation by following a 
population of large planetary embryos 
with a numerical model based on Chambers' Mercury code \citep{chmb02} 
which takes into account all mutual gravitational forces among 
the embryos as well as mutual collisions. Our main concern is here 
to see if mutual embryos accretion can lead to a final planet at the 
right place, i.e. $\simeq 2.1\,$AU, and within $10\,^{7}\,$years, i.e. the 
typical survival time for circumstellar discs \citep[e.g.][]{arm03}. 
 
The dynamical conditions for accretion within a 
planetesimal swarm perturbed by the companion of \gc\ and affected by 
gas drag are investigated in section 2. 
Section 3 is devoted to the study 
of the final accumulation of massive embryos into a core. 
In Section 4 we discuss the results.

\section{Relative velocities among planetesimals} 
 
\begin{figure} 
%\makebox[\textwidth]{ 
%\includegraphics[width=\columnwidth]{gam1.ps} 
%\hfil 
\includegraphics[width=\columnwidth]{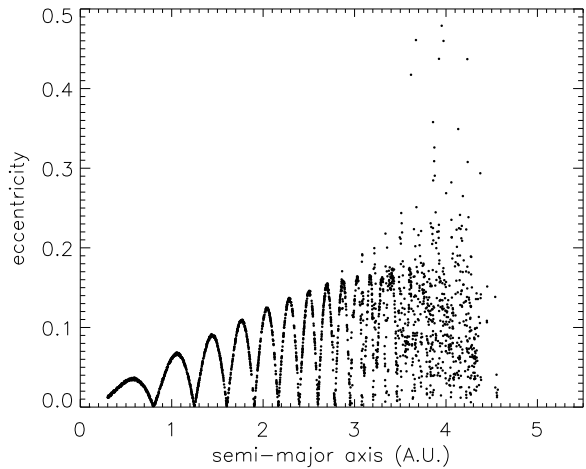} 
%} 
\caption[]{Distribution of particles eccentricity vs. 
semimajor axis  after $t=50000$\,years for a simulation 
where the gas drag is not included.} 
\label{gam1} 
\end{figure} 

Planetesimal accretion in a circumstellar disc occurs when 
mutual encounter velocities are lower than the 
escape velocity $V_{esc}$ of the mutually colliding bodies 
(corrected by a factor that accounts for the 
energy dissipation in the impact). In the absence of external perturbations, 
average eccentricities and inclinations in a planetesimal swarm are small, 
and relative velocities are always low enough to allow fast 
runaway accretion of larger bodies at timescales of the 
order of $10^4$--$10^5$ years \citep[e.g][]{lis93}. 
When the planetesimal population 
surrounds a star in a binary system, the gravitational pull of the 
companion star excites large eccentricities and the impact velocities 
may in some regions exceed by far the mutual escape velocity. In these regions 
the swarm would erode to dust rather than form a planet. 
 
A first step to test whether planetary formation is possible around \gc\ 
is thus to estimate the relative velocities between 
planetesimals orbiting within the dynamically stable region 
around the star. We have used a numerical code already adopted 
in similar studies \citep[see][and references therein]{the02}. 
This code computes the orbits of a 
swarm of massless particles under the influence of one (or 
several) gravitational perturbers. The particles can also be subjected 
to gas drag forces. During the orbital computations, the relative velocities 
during all the mutual encounters are recorded. These velocities approximate 
with a satisfying degree of precision the impact velocities during 
planetesimal collisions. 
Their distribution tells us whether planetesimals accrete into 
larger bodies or erode into smaller pieces. 
 
In all our simulations we integrate the orbits of  2500 test 
particles initially 
distributed in the 0.3-5\,AU region. The initial eccentricities and 
inclinations satisfy the relation $i=e/2$ and are chosen such that 
the  average encounter 
velocity is $\langle \Delta v \rangle\simeq\,10\,$m.s$^{-1}$. 
We stop our simulations after 
$t_{final}=10^{5}$\,years, a typical timescale 
for the formation of planetary embryos \citep{lis93}.

\subsection{Gravitational model without gas drag} 
 
\begin{figure} 
\includegraphics[angle=0,origin=br,width=\columnwidth]{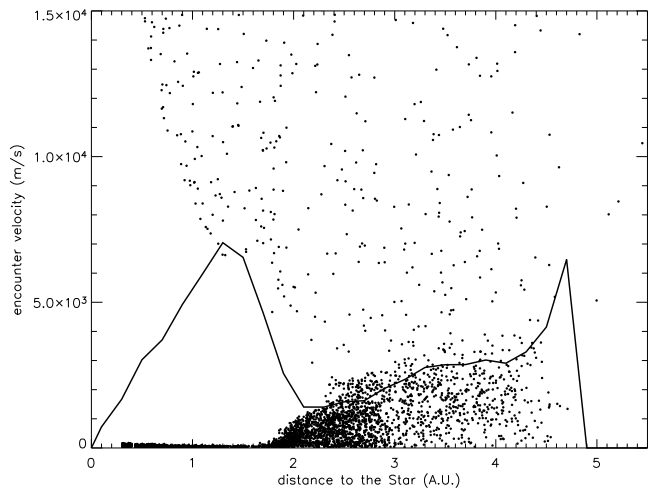} 
\caption[]{Plot of $\Delta$v vs. radial distance to the star 
of the planetesimal swarm at $t=50 000$ years. Each dot stands 
for an encounter that has occurred at a velocity $\Delta$v and at 
a distance $r$ from the star during 
$\lbrack$t--$10^{4}$,t+$10^{4}\rbrack$ . 
The continuous line is the average encounter velocity at 
each distance from the star.} 
\label{gam2} 
\end{figure} 
\begin{figure} 
\includegraphics[angle=0,origin=br,width=\columnwidth]{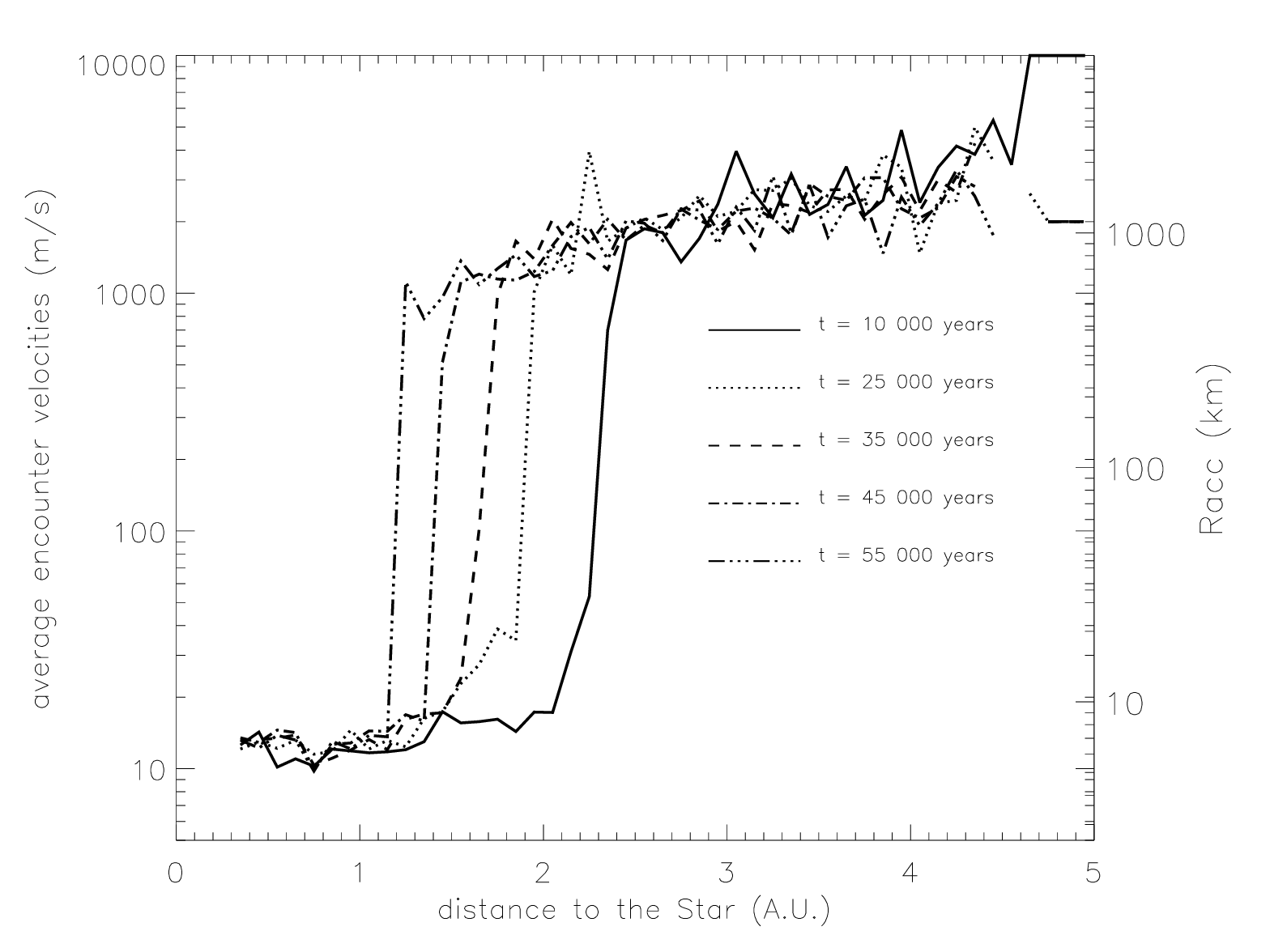} 
\caption[]{Evolution of the average encounter velocity 
distribution, at 5 different epochs, for the 
gravitational model without gas drag in a disc initially truncated at $4\,$AU. 
$R_{acc}$ is the size of the smallest body that can accrete matter,
after a collision with an equally sized object, 
for a given encounter velocity $\Delta v$. It is given by 
$R_{acc}= f \left( \frac{3} {8\pi \rho_{p} G} \right)^{0.5}\Delta v$, 
where $\rho_{p}=3\,$g.cm$^{-3}$ is the density of the body and 
$f$ is a coefficient accounting for the energy dissipation in an impact. 
We adopt here the 
usual assumption that 90\% of the energy is dissipated in the radial 
direction (e.g. Petit and Farinella 1993) which leads to 
$\langle f \rangle \simeq 0.7$ 
} 
\label{gam3} 
\end{figure} 

We performed a  first test simulation where only 
the gravitational forces of the two stars are accounted for. 
Fig.\ref{gam1} shows the major secular perturbations of the 
companion star on the planetesimal population. At the beginning 
of the simulation eccentricities progressively increase but the 
periastra of these forced orbits are almost aligned. 
As time goes on, however, 
there is a progressive dephasing process due to the different 
values of semimajor 
axis of the planetesimals in the swarm. This dephasing leads to the wavy 
pattern observed in  Fig.\ref{gam1}. The planetesimals 
orbiting close to the outer limit of the disc 
at $\simeq 4.5 \,$AU 
are more  strongly perturbed by the secondary star and they reach very 
high eccentricities. They also survive over $10^{5}$\,years 
and they cross the orbits of inner planetesimals in the 0--3\,AU region 
at high $\Delta$v (hence the wide "wing" of high 
$\Delta$v that is superimposed on the main swarm of lower relative velocities 
in Fig.\ref{gam2}). 
The contribution of these scattered planetesimals is critical since 
they increase relative velocities far beyond the accumulation threshold 
everywhere in the inner disc. 
However, it is a matter of debate whether planetesimals could indeed form 
in the outer edge of the stability region around the primary star. 
Moreover, theoretical calculations of binary--disc interactions predict that 
companions might truncate circumstellar discs at an outer radius of 
0.2-0.5 times the binary semimajor axes \citep{art94}. 
The grain coagulation process and the first impacts between the 
proto--planetesimals would in any case have been highly energetic, 
because of the companion's perturbations, possibly halting 
any further growth. For this reason, we performed an additional 
simulation with the planetesimal disc truncated beyond 4 AU. 
In Fig.\ref{gam3} we show the distribution of the impact 
velocities at different evolutionary times in this truncated disc. 
At the beginning, the secular oscillations do not induce a significant 
increase of $\Delta$v among planetesimals because of the strong phasing 
of all neighbouring orbits. Nevertheless, the oscillations 
get narrower with time and $e$ and periastron $\omega$ gradients between 
adjacent regions keep increasing. At some point, the orbital phasing is no 
longer strong enough to prevent orbital crossing 
between bodies with different semimajor axis. 
The inner limit for the region where orbit crossing occurs is very sharp, 
with $\langle \Delta v \rangle$ increasing from 10\,m.s$^{-1}$ to 
$\simeq1$\,km.s$^{-1}$ in less than 0.2\,AU (Fig.\ref{gam3}). 
Furthermore, this limit evolves inwards 
with time, so that the 2.1\,AU region is reached in less than 
$1.5\times10^{4}$\,years (Fig.\ref{gam3}). 
As a consequence, the timespan during which accretion 
of km-sized objects is possible within 2.1\,AU from the star, 
assuming that the swarm was 
initially aligned, is  typically of the order of $10^{4}$\,years. 
Numerical accretion models predict that this time span is 
in principle enough 
to allow the formation of 100 to 1000\,km-sized bodies 
\citep[e.g.][]{wet93,bar93,wei97}. However, the large relative velocities 
that build up after $10^{4}$\,years might halt any further growth 
of the accreted objects. It is even  
possible that the planetesimal formation process from the  
dust of the circumstellar disc does not necessarily lead to  
orbits that are initially aligned. In this case, the planetesimals  
would have from the beginning of their evolution high relative  
velocities. This might be the case if  
we suppose that the binary forms by direct stellar-like gravitational 
instabilities. The companion 
star would reach its present mass well before the onset of planetesimal 
accretion in the inner disc. 
Thus, the formation of large embryos in the 2\,AU 
region probably requires the presence of some additional mechanism.

\subsection{The effects of gas drag} 
\begin{table} 
\caption[]{Initial parameters for the reference gas drag runs} 
\label{param} 
\begin{tabular}{ll} 
\hline 
Number of bodies & 2500\\ 
Physical radius of bodies & variable (see graphs)\\ 
Initial semi-major axis & 0.3--5\,AU\\ 
Initial eccentricities & $0<e<10^{-3}$\\ 
Initial inclinations & $0<i<5.10^{-4}$\,rd\\ 
Gas density at 2 AU & $\rho_{g0}=2\,10^{-9}$g.cm$^{-3}$\\ 
Gas density radial profile & $\rho_{g}\propto\,r^{-2.75}$\\ 
\hline 
\end{tabular} 
\end{table} 

Frictional drag by the gas of the protoplanetary disc is an important 
factor in early planetary formation. 
It affects planetesimal orbits in two ways: 
 
\begin{itemize} 
\item It restores the periastron alignment \citep{mascho00}, 
preventing orbital crossing of orbits with different semimajor 
axes. At the same time, it partially damps the amplitude of 
oscillations in eccentricity 
induced by the companion star. 
 
\item It causes a drift towards the central star that is 
particularly fast for planetesimals in binary star systems: 
The forced component in the planetesimal eccentricity is 
large, in spite of the damp effect, and it causes a fast drift 
towards the star. 
\end{itemize} 
 
Following \citet{WD85}, we model the gas drag force as: 
\begin{eqnarray} 
        \vec{F} & = & - K v \vec{v} 
\end{eqnarray} 
where $\vec{F}$ is the force per unit mass, $\vec{v}$ the velocity 
of the planetesimal with respect to the gas, $v$ the velocity modulus, 
and $K$ is the drag parameter. It is a function of the physical 
parameter of the system and is defined as: 
\begin{eqnarray} 
        K & = & \frac{3 \rho_g C_d} {8 \rho_{pl} R_{pl}} 
\end{eqnarray} 
where $\rho_g$ is the gas density, $\rho_{pl}$ and $R_{pl}$ the 
planetesimal density and radius, respectively. $C_d$ is a 
dimensionless coefficient related to the shape of the body 
($\simeq 0.4$ for spherical bodies). 
As it appears from the expression of the 
parameter $K$, the relevance of the drag 
force on the dynamics of the planetesimals 
depends on the ratio between the gas density 
and the planetesimal size. As a 
consequence, the results we obtain for a particular value of $\rho_{g}$ 
and $R_{pl}$ can be extended to different combinations of the 
two parameters without any additional simulation. 
 
We take as a reference 
value for the gas density that of \citet{bod00} who 
modeled the {\it in--situ} formation of a giant planet around 
47\,UMa at 2\,AU. Of course, the two systems 
are not exactly comparable, the planet around 47UMa being 
$\simeq50\%$ more 
massive whereas the star is $\simeq60\%$ less massive. While these two values 
might compensate one another it is difficult to quantitatively 
estimate how much exactly. However, the \citet{bod00} study 
gives a good reference value for the in-situ formation of a giant planet 
on an inner orbit. 
According to these authors, this formation requires a 
local gas density $\rho_{g(2AU)}=2\,\times \,10^{-9}$g cm$^{-3}$, 
about an order of magnitude higher than the value deduced from the 
\citet{haya81} minimum mass solar nebula. 
We here use this value for $\rho_{g(2AU)}$ 
and adopt the classical $\rho_{g}\propto\,r^{-2.75}$  radial profile 
of Hayashi (1981). 
Different values of $R_{pl}$ are taken in independent simulations 
to explore the parameter space. 
The initial parameters adopted for our reference runs 
are summarized in Tab.\ref{param}. 

\begin{figure} 
\includegraphics[angle=0,origin=br,width=\columnwidth]{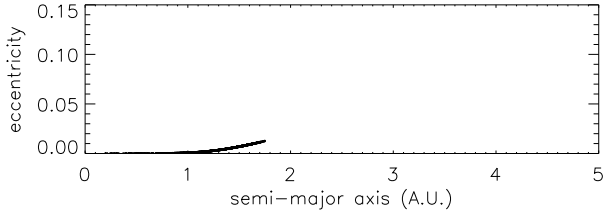} 
\caption[]{
eccentricities vs. semi-major axis at 
 $t=2500$\,years for 200m-sized planetesimals. The gas density 
is taken from \citet{bod00} with a nominal 
Hayashi (1981) radial profile.} 
\label{gam4} 
\end{figure} 
\begin{figure} 
\includegraphics[angle=0,origin=br,width=\columnwidth]{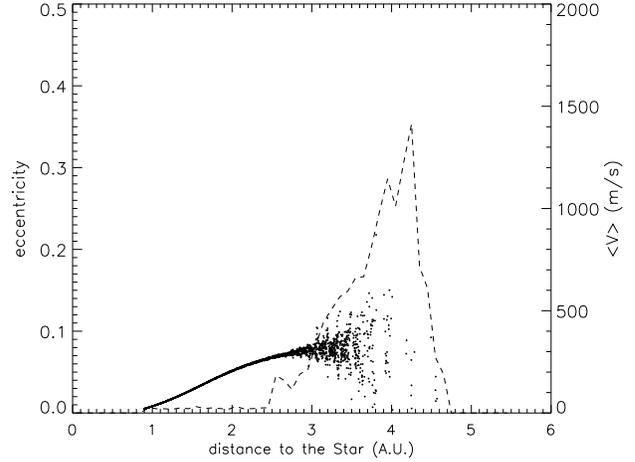} 
\caption[]{Eccentricity vs. semimajor axis plot 
after $t=50000$\,years for 10\,km-sized planetesimals. 
The dashed line is the corresponding average encounter velocity distribution 
as a function of the distance to the star. 
Gas density is the same as in Fig.\ref{gam4}.
} 
\label{gam5} 
\end{figure} 
\begin{figure} 
\includegraphics[angle=0,origin=br,width=\columnwidth]{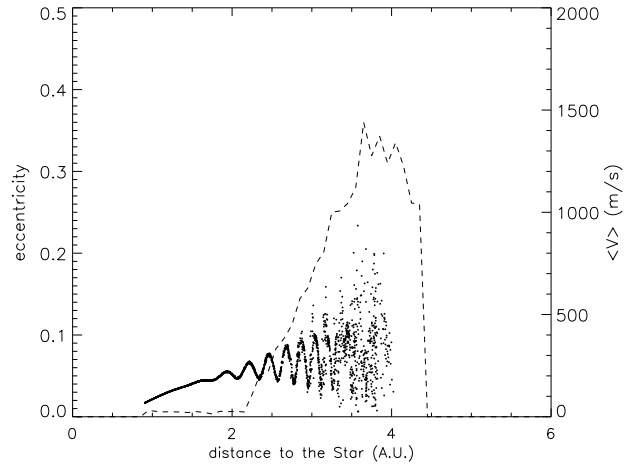} 
\caption[]{Same as Fig.\ref{gam5} but for 50\,km-sized bodies} 
\label{gam6} 
\end{figure}% 
 
\begin{figure} 
\includegraphics[angle=0,origin=br,width=\columnwidth]{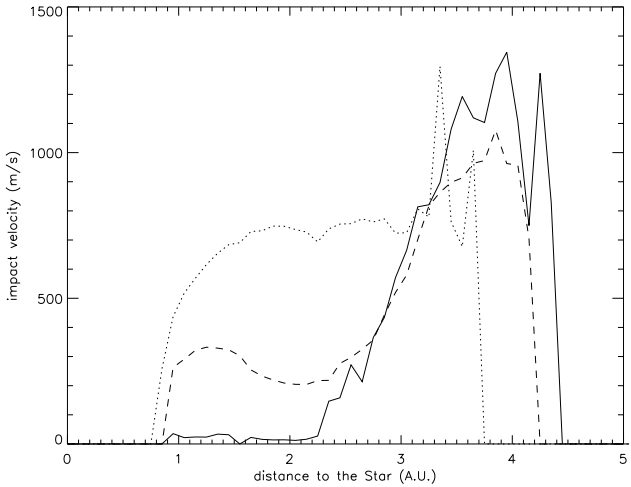} 
\caption[]{Mutual relative velocities, at $t=30000$\,years,
for a disc with 3 different 
populations of objects: 1\,km, 10\,km and 50\,km in sizes. The solid line 
stands for $\Delta v$ among 50\,km bodies, the dashed line 
for $\Delta v$ between 50\,km and 10\,km bodies and the dotted 
line for $\Delta v$ between 50\,km and 1\,km bodies.
The gas disc has the same characteristics as in Figs.\ref{gam4},\ref{gam5}
and \ref{gam6}
} 
\label{gam6b} 
\end{figure} 

\begin{figure} 
\includegraphics[angle=0,origin=br,width=\columnwidth]{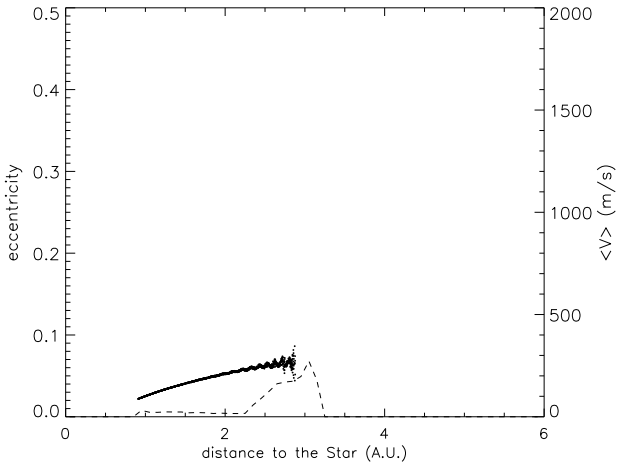} 
\caption[]{Same as Fig.\ref{gam6} (50\,km bodies at $t=50000$\,years)
for a planetesimal  disc initially truncated 
at 3\,AU. The gas density decreases radially as  $r^{-1.5}$.} 
\label{gam7} 
\end{figure} 
For planetesimal accretion to occur, 
a balance between different competing mechanisms 
is needed. For too small planetesimals, there is the 
risk that a too strong gaseous friction  leads to a 
fast inward drift that prevents accretion of larger 
bodies far from the star. 
This is illustrated in Fig.\ref{gam4}: a 
population of 200m-sized bodies migrates towards the 
star in less than  $2 \times 10^{3}\,$years leaving the region 
beyond 2\,AU totally depleted of material. 
On the other hand, large planetesimals, that are less 
affected by gas drag, do not significantly migrate 
but their periastron alignment is weaker. 
If it gets too weak, then 
encounter velocities may not be reduced below the value that 
allows accretion into larger bodies. 
 
In between these two accretion inhibiting cases, there is a planetesimal 
size-range where periastron alignment and eccentricity damping are efficient 
enough to reach very low $\Delta$v without  removing bodies 
from the system on a short timescale. As an example, for 10\,km bodies 
the periastron alignment keeps the 
impact velocities below 10m/s within 2.5 AU from the star (Fig.\ref{gam5}) 
and the drift rate is also slow enough not to deplete this 
region of the disc within $10^5\,$years. Fig.\ref{gam6} 
shows the limiting case, i.e. the run with the biggest $R_{pl}$ 
(50\,km) in which accretion is still possible at 2.1\,AU. 
Beyond 2.2 AU the collision velocities begin to grow quickly as the 
secular perturbations are dephased. 

As the planetesimal accretion proceeds, larger bodies are formed and  
they still collide with the smaller ones that possibly still  
make up most of the  
mass of the swarm. It is relevant to verify whether the impact velocities  
between large and small planetesimals still favour accretion rather then 
fragmentation. The perihelion alignment is indeed different depending on  
the size of the body. Smaller bodies  tend to align their perihelia towards  
$270^{\circ}$ \citep{mascho00} while larger planetesimals,  
less affected by the drag force,  
align to larger values. At the orbital crossing the impact velocities may  
thus be higher compared to those of equal size bodies. We performed  
additional simulations where we include different size planetesimals.  
Fig.\ref{gam6b} shows the relative velocities between populations 
of different size planetesimals. For  
50 \,km target objects, colliding speeds significantly increase when the 
impactor sizes get smaller, especially in the otherwise low $\Delta v$ region 
below 2.5\,AU. The $\Delta v$ might reach $\simeq\,300$m.s$^{-1}$ when 
10\,km impactors are considered, and even  
$\simeq\,700$m.s$^{-1}$ for 1\,km objects. 
However, it can be shown that such impact velocities still lead to 
accretion. Adopting the collisional  
algorithm described in \citet{pet93}, we find that  a collision 
between a 50\,km and a 10\,km object at $300\,$m.s$^{-1}$ leads 
to net accretion of $\simeq\,85\,\%$ of the impactor's mass, and 
that a collision 
between a 50\,km and a 1\,km object at $700\,$m.s$^{-1}$ leads 
to net accretion of $\simeq\,95\,\%$ of the impactor's mass.

\begin{figure} 
\includegraphics[angle=0,origin=br,width=\columnwidth]{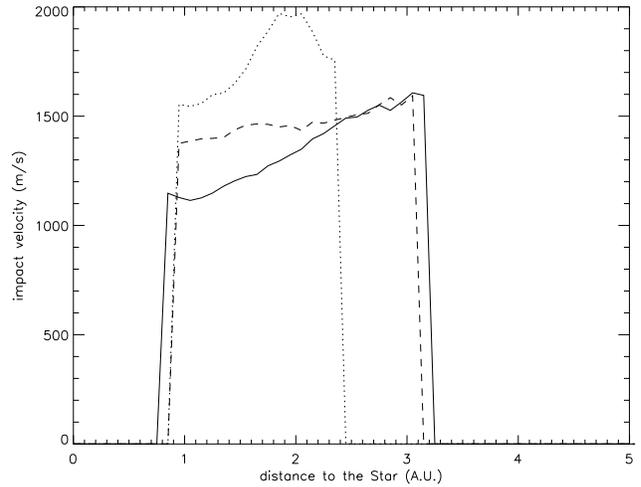} 
\caption[]{Impact velocities, at $t=15000$\,years,
on 500\,km bodies for 50\,km (continuous 
line), 10\,km (dased line) and 1\,km (dotted line) impactors. 
Same high density truncated disc as in Fig.\ref{gam7}. 
The absence of $\Delta$V values beyond 2.5 AU for 1\,km impactors 
is due to the fast removal of such small objects from these regions} 
\label{vit500} 
\end{figure} 

Conditions slightly more favourable to planetary formation, i.e.
allowing accretion of objects bigger than 50\,km, are met if 
we adopt a flatter gas density profile or if we truncate the 
disc closer to the star. In Fig.\ref{gam7} we show the relative 
velocity distribution in a disc of 50\,km radius planetesimals 
cut at 3\,AU from the central star and 
a gas density profile in $r^{-1.5}$ granting a 
stronger drag force between 2 and 3 AU. Additional simulations 
show that under these conditions low $\Delta$v 
are maintained for bodies up to $\simeq\,200$\,km, while the 
inner limit of the low $\Delta$v region never extends beyond 2.5 AU. 

For bodies in the 200--1000\,km range, higher $\Delta$v are obtained 
in the $a<2.5\,$AU region, especially for 
collisions with smaller planetesimals, which are probably the 
most frequent impactors on such bigger objects (Fig.\ref{vit500}). 
However, such high $\Delta$v impacts still lead to net 
mass accretion. Using again \citet{pet93} algorithm 
for velocities obtained in Fig.\ref{vit500} on a 500\,km target, one gets that 
95\% of the impactor's mass is reaccreted after a 1200\,m.s$^{-1}$ 
collision with a 50\,km impactor, 97\% for a 1400\,m.s$^{-1}$ impact 
with a 10\,km body and 98\% for a 1700\,m.s$^{-1}$ 
collision with a 1\,km impactor. 
So even considering the large uncertainties on collision and  
reaccretion physics, it seems that we are far below the accretion threshold 
for 500\,km bodies in our "extreme" gas disc case,
i.e. high gas density in the outer regions (flat radial distribution)
and cut-off at 3\,AU for the planetesimal swarm. 
 
Furthermore, even with lower gas densities, the final accretion 
phase of objects in the 200-1000\,km range can be favoured by  
additional mechanisms that cannot be accounted for  
in our N--body simulations. 
Dynamical friction produced by the gravitational interactions  
between planetesimals tends to produce equipartition of  
energy \citep{wet89}. As a consequence, larger bodies have lower  
random velocities favouring softer impacts. Self--gravity of  
planetesimals may also restore periastron alignment, as recently  
showed by Kokubo (personal communication). 
The scenario that  
comes out from our simulations can thus be summarized as following:  
Small to medium size planetesimals are strongly affected   
by gas drag and their relative velocities are kept low by the  
periastron alignment. When the gas drag weakens, additional  
dynamical mechanisms, such as dynamical friction and self--gravity, 
are still active and help large planetesimals to continue their  
growth. In case of massive protoplanetary discs, gas drag can be able  
to do all the job, allowing the growth of planetesimals until they  
are big enough to even sustain high velocity impacts. 
 
It should also be noted that the 
high gas densities required to maintain the periastron alignment up to  
large size planetesimals might not be unrealistic in the case of the  
\gc\ system. It is reasonable to expect that accretion discs surrounding 
F2 stars, progenitor of K III giant stars as \gc\, are massive compared 
to the minimum mass solar nebula. Even the density value for the gas  
adopted by \citep{bod00} for the disc surrounding 47 UMa, a G0V star,  
may be a lower limit for discs around F2 stars.    
 
An additional conundrum for planetary formation in \gc\ is 
whether planetesimals could form in the outer regions of the 
disc perturbed by the gravity of a secondary star.  How 
sensitive was the dust sticking mechanism to the 
gravitational pull of the star?  It seems reasonable to assume 
that the sticking mechanism was more  efficient 
in the inner disc, where the perturbations of 
the secondary star were weaker. The truncation of the 
planetesimal disc beyond a few AU from the star might thus be supported 
by the physics of planetesimal formation. At which radial distance 
did the truncation in the planetesimal population become significant? 
Only detailed models of planetesimal formation that include the 
gravitational effects of the companion star can answer this 
question. 
 
On the other hand, we should not forget that for too high gas 
densities, small bodies would spiral towards the star on 
a short timescale. The question is then: 
did accretion proceed fast enough to prevent such 
a loss of small bodies where a large fraction of the 
mass lies? 
Planetesimal accretion in the \gc\ system thus requires 
a delicate balance between perihelia alignment at large sizes, fast 
spiralling for small bodies, and proper values of the impact probability. 
 
A numerical model that includes all these effects at once 
is, at present, beyond computer capabilities. 
The main result of this section is that 
gaseous friction opens a window for starting planetesimal accretion 
within 2.5 AU from the star. 
 
\section{Core formation by accretion of protoplanets} 
 
In this section, we shall assume that planetesimal accretion could 
take place and lead to the formation of planetary embryos. 
This embryo formation could have followed the scenario described 
by \citet{kor01}. These authors have modeled planetesimal accretion 
in systems with a massive external perturber showing 
that the combination 
of gas drag, collisions, and secular perturbations of a massive external 
body favours orderly growth in the initial phases of planetary 
accretion followed by a phase of ''type II'' runaway growth. 
 
We here investigate whether giant impacts 
between such massive embryos in \gc\ can lead to the formation 
of a planetary core whose orbit resembles that of the 
observed planet and whose mass is at least $10\,M_{\oplus}$, 
the mass required to trigger the final phase where the core 
accretes the remaining gas through rapid infall and forms 
a giant planet \citep{pol96}. 
We also evaluate how the timescale of 
the core formation depends on the initial mass of the 
embryos, on their number, and on their spatial distribution. 
 
We have simulated, within a full 
N--body model, the evolution of a population of  protoplanets 
into a massive core by integrating the orbits of 
a swarm of planetary embryos 
distributed between 1 and 2.5 AU where, according to the 
previous computations, planetesimal accretion is mostly efficient. 
The orbital evolution of the embryos and their collisions 
are computed with Chambers' 
Mercury code \citep{chmb02} which we modified to account for the 
large perturbations of the binary companion. The number 
and initial masses of the embryos are derived assuming a reference 
surface density of solid material at 2.1 AU ranging from 
50 to 100\,g cm$^{-2}$. 
As in the previous section, these values are derived from the 
\citet{bod00} estimates for formation of the planetary companion of 47 UMa. 
The embryos initial masses range from a few 
Lunar mass to the mass of Mars, 
depending on the adopted value of the disc density. 
In all the simulations we assume that only a fraction of the mass 
of the solids in the disc has accumulated into planetary embryos 
while the remaining mass is still in smaller planetesimals. 
This percentage varies from 50\% to 75\% of the total mass 
in different models. We also adopted three different 
distributions of the protoplanets as a function of the 
distance from the star. 
This is motivated by the  
uncertainty about the details of the planetesimal accretion process. 
Runaway growth, type II runaway growth, orderly growth,  
or even oligarchic growth can occur depending on the delicate  
balance between the mass and the velocity distribution. 
\citet{kor01} have shown that, for a particular binary system, type II  
runaway growth is to be preferred. However, the 
type of growth strongly depends on the binary orbital and physical  
properties, and on the disc parameters like the  
density and mass distribution.  
We prefer to cover a large spectrum of possibilities by assuming in our  
simulations different but equally possible initial  
conditions for the protoplanets: 
 
1) A population of embryos whose number and location is computed 
according to a superficial density $\sigma$ constant on average 
between 1 and 2.5 AU. 2) 
A $\sigma$ that decreases as $ 1 / r$ giving a constant 
mass for the embryos 
populating annular portions of the disc surrounding the star. 
3) A fixed initial radial distance between the planetary 
embryos expressed in Hill's radii. 
In some of the simulations we even include a ''proto--core'', 
an embryo with an Earth  mass located at 2.1 AU. 
Initially, all embryos have eccentricities 
lower than 0.04 and inclinations lower than $1^{\circ}$ with 
respect to the orbital plane of the binary system. 
 
\begin{figure} 
\includegraphics[angle=-90,origin=br,width=\columnwidth]{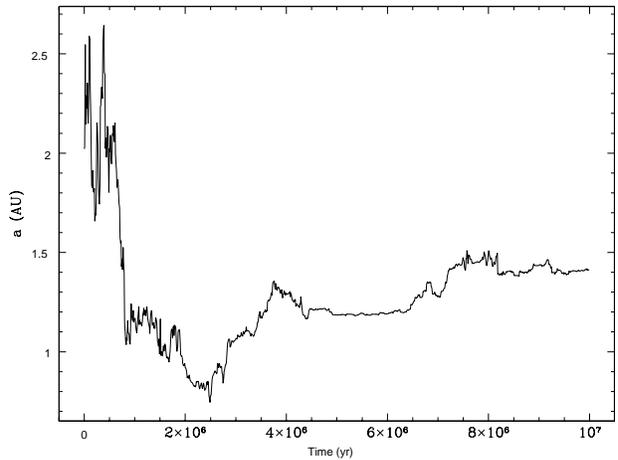} 
\caption[]{Migration of the growing core due to scattering of protoplanets. The 
initial disc extends from 1.5 to 2.5 AU and has a total mass of 
75\,$M_{\oplus}$.} 
\label{gam8} 
\end{figure}

\begin{figure} 
\includegraphics[angle=-90,origin=br,width=\columnwidth]{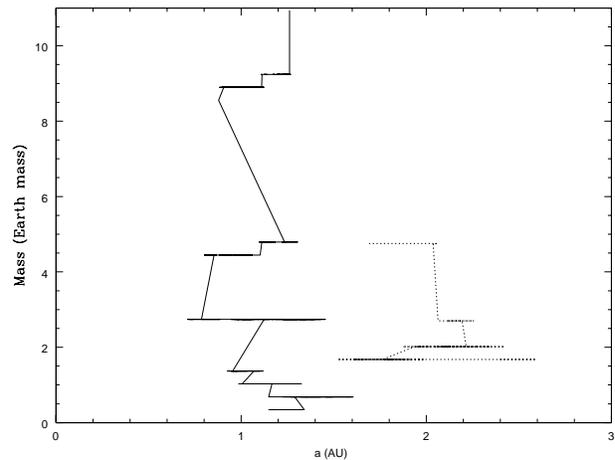} 
\caption[]{Evolution, in the (M,$a$) plane, of a $1\,M_{\oplus}$ proto-core 
initially put in the outer disc and of the largest embryo growing in 
the inner region. The rapid growth of the inner embryo results 
in the capture of the outer proto-core (see text for details).} 
\label{gam9} 
\end{figure}

Our simulations with an initial total embryo mass 
of 25 $M_{\oplus}$, about 50\% of the solid mass 
in a disc with $\sigma = 50$ g.cm$^{-2}$, 
fail to create a core of 10$M_{\oplus}$ within 10Myr, 
the typical lifetime of a gaseous disc. 
The maximum mass of the core achieved in these simulations 
is 6$\,M_{\oplus}$, which 
might not be enough to trigger the final gas accretion phase. 
If we increase the total embryo mass to 
35$\,M_{\oplus}$, a core of 8--10$\,M_{\oplus}$ 
can form in a few cases. All simulations with an initial mass ranging from 
50 to 75 $M_{\oplus}$, 
compatible with $\sigma \simeq$ 100 g~cm$^{-2}$, 
lead to the formation of a core with 
a mass up to 20 $M_{\oplus}$ within 10 Myr. 
However, our simulations all show that the core always ends up 
within 1.5 AU from the primary star, while the observed 
planet is at about 2.1 AU. Even when including in the 
initial protoplanet population a bigger ''proto--core'' of 
1 Earth mass at 2.1 AU, the final accreted core always ends up 
between 1 and 1.5 AU.  
Two distinct mechanisms might account for this outcome: 
1) In some cases a core might begin to form in the outer regions 
but it migrates inwards 
due to planetesimal scattering reinforced by the 
gravitational forces exerted by the binary star which excites the embryos' 
eccentricities (Fig.\ref{gam8}). 
2) In some other cases a core begins to form at 
around 1 AU. It grows at a faster rate because 
of the shorter Keplerian orbital period of bodies in inner 
orbits. This core is able to accrete large protoplanets in outer orbits, 
since the orbital eccentricity is large due to 
the companion's perturbations, and the final result is a core around 
1.5 AU and almost no remaining material beyond this position. This result 
hold even when a large ``proto-core'' is initially placed in the outer 
disc (Fig.\ref{gam9}). 
These two mechanisms are peculiar of 
protoplanetary accretion in binary star systems, where large 
eccentricities among the embryos are excited by the 
gravitational pull of the secondary star.  
In a few simulations the 'proto--core'  is ejected out of the system 
before it reaches 10$\,M_{\oplus}$ , since it is close 
to the border of the stability region. 
 
\begin{figure} 
\hskip -1 truecm 
\includegraphics[angle=-90,origin=br,width=14cm]{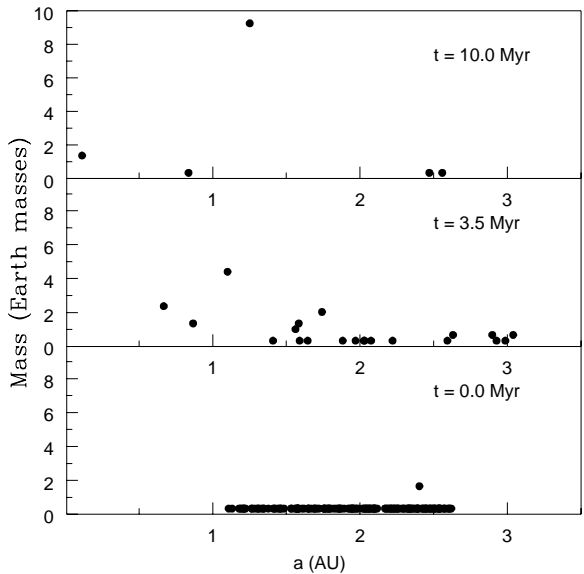} 
\caption[]{Snapshots of protoplanet accretion in the \gc\ 
 system. The initial mass in the embryos is 
75 $M_{\oplus}$ and the superficial density $\sigma$ 
decreases with the radial distance as 
$1/r$. 
} 
\label{gam10} 
\end{figure}

In Fig.\ref{gam10} 
we show the outcome of a simulation with an intermediate 
value for the total embryo  mass (75 $M_{\oplus}$) 
and a proto--core 
initially located at 2.3 AU. The 
proto--core grows faster than nearby protoplanets 
but it migrates inward due to the scattering of the other 
bodies. It settles at about 1.4 AU and its 
mass reaches almost 10  $M_{\oplus}$. The final giant planet 
is then expected to orbit closer to the primary star than 
the observed planet in \gc\ 


\section{Discussion and conclusions} 
 
The previous results show that planetesimal accretion in 
\gc\ depends on a delicate balancing between gas drag and 
secular perturbations by the secondary star. If this balancing 
is met, then type II runaway growth \citep{kor01} might possibly 
lead to the formation of massive planetary embryos within 2.5 AU 
from the central star.  In the subsequent phase when 
giant impacts between the embryos build up a massive 
core, the major problem  is not the timescale 
but the final location of the planet, which is always well inside 
the actual position of the observed planet. 
Several explanations might be proposed to account for 
this discrepancy. 
A first hypothesis involves the evolution of the whole system. 
It is possible that the distance between the two 
stars was larger when 
the planet formed and that after the planet formation 
additional mechanisms pushed the secondary star on 
an inner orbit. This could be the case if the \gc\ system 
was born in a clustered 
environment, where close encounters with other young stars may cause 
perturbations of the binary orbit that tend to shrink it \citep{heg85}. 
A different way to reduce the orbit of a binary system is related 
to the possibility that, originally, the system was triple or more. 
The ejection of one or more stars causes a transfer of binding energy 
and an eventual reduction of the binary separation \citep{rei00}. 
A more complex mechanism is related to the formation 
of more than one giant planet around \gc\ . 
If the circumstellar disc around the main star was significantly 
more massive, with a superficial density of solids higher 
than 100 g cm$^{-2}$, it is possible that two or more giant planets 
formed around the main  star. 
Mutual scattering among these planets ejected one, two or more of them 
out of the system leaving a single planet in the observed orbit 
\citep{gra04}. 
Another and more radical solution would be to renounce 
the core-accretion model in favour of the alternative 
disc instability scenario \citep{bos01}, but this scenario 
remains to be quantitatively tested for the \gc\ system. 

{} 
\clearpage 
 
\end{document}